\documentclass[paper,a4paper,12pt]{JHEP3}
\usepackage{amssymb,graphicx}

\newcommand{\non}{\nonumber \\}

     \newcommand{\bet}{\beta}

\newcommand{\zet}{\zeta}      \renewcommand{\th}{\theta}
 
     \newcommand{\lam}{\lambda}
   \newcommand{\sig}{\sigma}
 
   \newcommand{\ome}{\omega}

     \newcommand{\Lam}{\Lambda}

    \newcommand{\cN}{{\cal N}}
\newcommand{\cO}{{\cal O}}

\newcommand{\pa}{\partial}

\newcommand{\rar}{\rightarrow}

\newcommand{\gsim}{ \lower .75ex \hbox{$\sim$} \llap{\raise .27ex \hbox{$>$}} }
\newcommand{\lsim}{ \lower .75ex \hbox{$\sim$} \llap{\raise .27ex \hbox{$<$}} }
\newcommand{\EM}{\rm\scriptscriptstyle EM}

\renewcommand{\Im}{\mbox{Im}}

\title{Photon and Dilepton Production in Soft Wall AdS/QCD}
\author{A. Nata Atmaja${}^{1,2}$ and K.Schalm${}^{1,2}$\\
${}^1$Institute Lorentz for Theoretical Physics\thanks{Permanent address from September 1st 2007.}\\
Niels Bohrweg 2\\
2333CA Leiden, the Netherlands\\
\\
${}^2$Instituut voor Theoretische Fysica\\
Valckenierstraat 65\\
1018XE Amsterdam, the Netherlands\\
\email{ardian@lorentz.leidenuniv.nl, kschalm@lorentz.leidenuniv.nl}
}

\abstract{
We consider the Soft-Wall-model of AdS/QCD to calculate photon production
in strongly coupled Quark Gluon Plasma (sQGP). 
The IR cut-off only affects the low-frequency-component of the production rate.
The full spectral function is determined numerically and shows
remarkable similarity to computations of the photon production rate in
AdS-duals of $\cN=2$ theories with massive flavor. It is further support
that
Soft-Wall AdS-QCD correctly captures the IR physics of the chiral
perturbation theory regime of QCD. We confirm this by relating 
the IR-effects of the massive flavor deformations to the AdS/QCD soft
wall cut-off. The AdS/QCD spectral function is smooth, however, and
unlike massive flavor models shows no spectral peaks.}


\preprint{ITFA-2008-02}

\begin{document}

\section{Introduction}
One of the current challenges in theoretical particle physics is to
compute properties of the strongly coupled sQGP discovered at
RHIC. AdS/CFT tools have given us some insight into the strongly
coupled thermodynamics of gauge theories
\cite{maldacena97,Gubser:1998bc,witten98-1,Gubser:1996de,Klebanov:1996un,witten98-2,policastro02}. However, it
remains a mystery why these, mostly $\cN=4$ supersymmetric, YM calculations work well for QCD. Part of the challenge is to either understand why this is so, or to find AdS duals of theories resembling QCD closer than $\cN=4$ SYM. In this latter context a phenomenological AdS dual to Chiral perturbation theory constructed by Erlich et.al. is perhaps a good candidate \cite{erlich05}. It predicts the meson spectrum within 10\%-15\% accuracy; and with a smooth instead of a hard IR-cutoff predicts the critical $T_c$ surprisingly well~\cite{karch06,herzog06}.\footnote{The prediction of the critical deconfinement temperature in soft-wall model is really quite surprising; The neglection of backreaction invalidates the thermodynamic relations that are used to derived it. Detailed studies that include the effects of backreaction are e.g. \cite{Gursoy:2008bu,Gursoy:2008za,Gursoy:2009jd} for the backgrounds which are believed to correspond to large $N_c$ pure Yang-Mills theory and \cite{Gubser:2008ny,Gubser:2008yx} for the backgrounds that mimic QCD equation of state.}

This IR-cutoff is the essential new ingredient in AdS/QCD compared to
AdS/CFT. Here we shall investigate the effects of this cutoff on
photon and dilepton production rates at strongly coupling. Remarkably
the $\cN=4$ SYM CFT computation of these production rates suggested
they are not affected by a hard IR-cut-off even for temperatures
infinitesimally above the cut-off \cite{huot06}. Intuitively this
seems rather strange. At energies and temperatures close the QCD scale
IR effects should start to affect the production rate. We shall find
that for smoothly IR-cutoff AdS/QCD this is indeed the case. The
robustness of our phenomenological result of how photon production
rates are effected by changing the IR-cutoff is confirmed by a
calculation by Mateos and Pati\~{n}o \cite{mateos07} of the photon production
rate in AdS dual of a $\cN=2$ theory with massive flavor. Here the
flavor sector acts as the effective IR-cutoff, and we will be able to
show this by relating the mass-parameter to the Soft-Wall cut-off
scale. Soft-wall AdS/QCD is more crude than massive flavor models, of
course, and this is evident in the lack of spectral peaks that we
shall find.

Photon production in a medium such as sQGP was discussed in detail
both from strong and weak coupling point view in \cite{huot06}. We
briefly review this in section 2 and show there how the strong
coupling calculation is modified by considering AdS/QCD instead of
pure $\cN=4$ SYM. In section 3, we present our solution and discuss
its results in section 4 with a comparison to photon production in AdS
duals
of $\cN=2$ massive flavor theories. 

\section{Photon and dilepton production}

One of the observational phenomena in RHIC is the spontaneous
production of photons from the sQGP of hot charged particles. This
direct photon spectrum ought to be a good probe of the strongly
coupled quark-gluon soup, 
as the weakly 
interacting photons should escape nearly unaffected from the small finite size collision area~\cite{stan05}. 

As is described in \cite{huot06}, we can therefore regard the
dynamically formed sQGP to
first approximation as a field theory at finite temperature. 
For a standard perturbative electromagnetic current coupling $eJ^{\EM}_\mu A^\mu$, the first order photon production rate is then given by \cite{huot06,bell96}
\begin{equation}
\label{eq:2}
	d\Gamma_\gamma=\frac{d^3k}{(2\pi)^3 2 k^{0}} e^2 n_B(k^{0}) \eta^{\mu\nu}\left.\chi_{\mu\nu}(K)\right|_{k^0=|\vec{k}|}.
\end{equation}
Here $K\equiv(k^0,\vec{k})$ is a momentum 4-vector, $n_B(k^0)=1/(e^{\beta k^0}-1)$ the Bose-Einstein distribution function, and the spectral density
$\chi_{\mu\nu}(K)$ is proportional to the imaginary part of the (finite temperature) retarded current-current correlation function 
\begin{eqnarray}
\label{eq:22}
	\chi_{\mu\nu}(K)&=&-2\mbox{~Im}(G^{R,
\bet}_{\mu\nu}(K)) ,\non
G^{R,\bet}_{\mu\nu} (K) &=& \int d^4X e^{-iK\cdot X} \langle J^{\EM}_{\mu}(0)J_{\nu}^{\EM}(X) \rangle_{\beta} \th(-x^0)~.
\end{eqnarray}
At finite temperature, Lorentz invariance is broken by the heat bath. We can use the remaining rotational symmetry plus gauge invariance to simplify the retarded correlator to
\begin{equation}
\label{eq:15}
 G^{R,\bet\neq 0}_{\mu\nu}(K)=P^T_{\mu\nu}(K)\Pi^T(K)+P^L_{\mu\nu}(K)\Pi^L(K),
\end{equation}
Here the transverse and longitudinal projectors are $P^T_{00}(K)=0$, $P^T_{0i}(K)=0$, {$P^T_{ij}(K)=\delta_{ij}-k_ik_j/|\vec{k}|^2$}, and $P^L_{\mu\nu}(K)=P_{\mu\nu}(K)-P^T_{\mu\nu}(K)$, with $i,j=x,y,z$. 
We can trivially consider charged lepton production as well by
considering non-lightlike momenta for off-shell photons: The leptons
then result from virtual photon decay. Lepton pair production for each
lepton species in the leading order of the electromagnetic couplings
$e$ and $e_l$, is given by \cite{huot06,bell96}
\begin{equation}
\label{eq:1}	
d\Gamma_{l\bar{l}}=\frac{d^4K}{(2\pi)^4}\frac{e^2e^2_l}{6\pi|K|^{5}}[-K^2-4m^2]^{1/2}(-K^2+2m^2)n_b(k^0)\chi_{\mu}^{\mu}(K)\theta(k^0)\theta(-K^2-4m^2),
\end{equation}
with $e_l$ the electric charge of the lepton, $m$ the 
lepton mass, $\theta(x)$ a unit step function, and the spectral density $\chi_{\mu\nu}(K)$ is evaluated at the timelike momentum of the emitted particle pair.
Note that both $\Pi^T$ and $\Pi^L$ contribute to the dilepton rate,
but only $\Pi^T$ contributes to the photon emission rate, because the
longitudinal part must vanish for lightlike momenta, i.e. the unphysical longitudinal mode is not a propagating degree of freedom. 

Finally, fluctuation-dissipation relates the zero-frequency limit of the spectral density to the electrical conductivity $\sigma$:
\begin{equation}
\label{Kubo formula1}
	\sigma=\lim_{k^0\to 0}\frac{e^2}{6T}n_{B}(k_0)\eta^{\mu\nu}\chi_{\mu\nu}(k^0,\vec{k}=0),
\end{equation}
or, if $k_{\mu}$ is lightlike
\begin{equation}
\label{Kubo formula2}
	\sigma=\lim_{k^0\to 0}\frac{e^2}{4T}n_B(k_0)\eta^{\mu\nu}\left.\chi_{\mu\nu}(K)\right|_{|\vec{k}|=k^0}.
\end{equation}

\subsection{Photon and dilepton rates at strong coupling}
The AdS/CFT dictionary gives that the large $N_c$ limit of 
strongly coupled $d=4~\cN=4$ SYM theory at finite temperature $T$ has a dual
description in terms of five dimensional AdS-supergravity 
in the background of a black hole \cite{witten98-2}
\begin{equation}
\label{AdS black hole}
	ds^2=\frac{(\pi TR)^2}{u}\left[-f(u)dt^2+dx^2+dy^2+dz^2\right]+\frac{R^2}{4u^2f(u)}du^2.
\end{equation}
Here $f(u)=1-u^2$, with $u\in [0,1]$ a dimensionless radial AdS coordinate related through $u=(\pi T z)^2$ to standard AdS coordinates, and $R$ is the curvature radius of the
AdS space.\footnote{
We will keep to Lorentzian signature throughout since we seek
information regarding the response of the thermal ensemble to small
perturbations. This requires the use of real-time Green's functions
\cite{son02}.}
The metric~(\ref{AdS black hole}) has a horizon at $u=1$ with Hawking temperature $T$ and a boundary at $u=0$.

Qualitatively the same is expected hold for other 4-dim field
theories. As a model for low energy QCD we shall take the AdS dual of
chiral perturbation theory. This AdS/QCD
consists of the fields $A^a_{L\mu},A^a_{R\mu},$ dual to the
$SU(N_f)_L\times SU(N_f)_R$ currents and a scalar $X$ dual to the
quark condensate in an AdS background which is cutoff at some finite distance $u=u_0$ \cite{erlich05}. 
To this we add an extra $U(1)$ field, $V_\mu$ dual to the electromagnetic current  $J_{\mu}^{\EM}$.\footnote{In chiral pertubation theory this $U(1)$ external field is
a subgroup of the local $SU(3)_L\times
  SU(3)_R$ \cite{scherer05}.
}
Recall 
that
$u_0$ corresponds to the 
introduction of the QCD-scale in the field theory:
 it enforces the mass-gap by hand by explicitly cutting-off any dynamics in the IR. For the reasons we explained in the introduction, here we are going
to use a soft wall cut-off \cite{karch06,herzog06}. Formally we can
introduce this
cut-off by modifying the AdS bulk action to (we give only the  term relevant for calculating the
photon production rate)
\begin{equation}
\label{AdS action}
 S \sim \int
 d^5x\sqrt{g}\left(-\frac{1}{4}F_{AB}F^{AB}+\cdots\right)
 ~~\Rightarrow~~ S \sim \int d^5x\sqrt{g}e^{-\Phi}\left(-\frac{1}{4}F_{AB}F^{AB}+\cdots\right).
\end{equation}
Here $A,B=t,x,y,z,u$ and the ``dilaton''  takes the fixed
form $\Phi=cu$ where $c={\Lambda^2_{IR}\over (\pi T)^2}$, with $\Lambda_{IR}$ the IR scale below which physics is cut-off. This introduction into
the action is formal in the sense that (1) we shall not consider $\Phi$ a
dynamical field and (2) we assume that the presence of
the cut-off does not affect the geometric AdS background, see also
\cite{herzog06}. We thus still work with the metric (\ref{AdS black
  hole}) for the finite temperature version of AdS/QCD, but with the
equation of motion for the fluctuations derived from action (\ref{AdS
  action}). We will discuss the validity of this approach in detail in
section \ref{sec:concl-soft-wall}.

For photon production, we need only the $U(1)$ gauge field equation of
motion  $\partial_A\left(\sqrt{g}e^{-cu}g^{AB}g^{CD}F_{BD}\right)=0$
with $F_{AB}=\partial_AV_B-\partial_BV_A$ the Maxwell field
strength. The 4d electric fields are $E_i\equiv F_{ti}$ with
$i=x,y,z$. Note that we use $A$ as a vector index and $V_{B}$ for the AdS
gauge field. To compute the AdS boundary 2-point correlation function
from which to extract the spectral density $\chi_{\mu\nu}$, we \mbox{follow
\cite{huot06}} and split the equation of motion into parts
perpendicular ($V_x,V_y\equiv V_{\perp}$) and parallel \mbox{($V_z\equiv
V_{\parallel}$)} to a predefined spatial three-momentum
$\vec{k}=(0,0,k)$, the Gauss constraint ($V_0$ e.o.m.) and the  radial AdS ($V_u$) equation of motion. 
After a Fourier transformation along $t,x,y,z$, and defining
$\omega=\frac{k^0}{2\pi T},~q=\frac{k}{2\pi T}$, 
we find respectively
\begin{equation}
\label{perpendicular EM}
	\partial_u^2V_\perp+\left(\frac{\partial_uf}{f}-c\right)\partial_uV_\perp+\frac{\omega^2-q^2f}{uf^2}V_\perp=0,
\end{equation}
\begin{equation}
\label{eom V_t}
	\frac{q}{uf}(qV_t+\omega V_\parallel)-\left(\partial_u^2V_t+i(2\pi T)\omega\partial_uV_u\right)+c\left(\partial_uV_t+i(2\pi T)\omega V_u\right)=0,
\end{equation}
\begin{equation}
\label{eom V_z}
	\frac{\omega}{uf^2}(qV_t+\omega V_\parallel)+\left[(\frac{\partial_uf}{f}-c)\partial_uV_\parallel+\partial_u^2V_\parallel\right]-i(2\pi T)q\left[(\frac{\partial_uf}{f}-c)V_u+\pa_uV_u\right]=0.
\end{equation}
The equation of motion for $V_u$, 
\begin{equation}
\sqrt{g}e^{-cu}g^{uu}\left(g^{tt}\partial_tF_{tu}+g^{\parallel\parallel}\partial_\parallel F_{\parallel u}\right)=0,
\end{equation}
can be simplified to
\begin{equation}
\label{V_u}
	V_u=\frac{i}{2\pi T}\frac{(\omega\partial_uV_t+qf\partial_uV_\parallel)}{(\omega^2-q^2f)}.
\end{equation}
Let us define $E_{\perp}=\ome V_{\perp}$ and $E_\parallel=qV_t+\omega
V_\parallel$. From Eq. (\ref{perpendicular EM}) and
combining eq. (\ref{eom V_z}) with eq. (\ref{eom V_t}) and eq. (\ref{V_u}) in the gauge $V_u=0$ we obtain the
two decoupled equations
\begin{eqnarray}
\label{perpendicular EM2}
\partial_u^2E_\perp+\left(\frac{\partial_uf}{f}-c\right)\partial_uE_\perp+\frac{\omega^2-q^2f}{uf^2}E_\perp=0,\\
\label{parallel EM}	\partial_u^2E_\parallel+\left[\frac{\omega^2\partial_uf}{f(\omega^2-q^2f)}-c\right]\partial_uE_\parallel+\frac{\omega^2-q^2f}{uf^2}E_\parallel=0.
\end{eqnarray}
We shall need to solve these two equations to obtain the spectral
density $\chi_{\mu\nu}$. These differential equations
(\ref{perpendicular EM2}) and (\ref{parallel EM}) have three regular
singular points at $u=\pm1,0,$ and one irregular singular point at
$\infty$.\footnote{Recall that an irregular singular point for a differential equation $y''+P(x)y'+Q(x)y = 0$ is a point $x_0$ for which either $\lim_{x\rar x_0} (x-x_0)P(x)$ or $\lim_{x\rar x_0} (x-x_0)^2Q(x)$ diverges. The point at infinity is irregular if $\lim_{x\rar \infty}(2- x P(x))$ or $\lim_{x \rar \infty} x^2Q(x)$ diverges. Using that $f=(1-u^2)$ one clearly sees how the introduction of $c$ introduces a divergence in $\lim_{u \rar \infty} 2-u(\pa_u\ln f-c) = \lim_{u\rar \infty} 2+2u^2/(1-u^2)+uc$. } 

Formal solutions for such equations are difficult to construct. Note that the irregular
nature of the point at infinity becomes regular when we remove the
IR-cutoff $c$. The irregular point, however, is outside the physical
region of interest $u \in (0,1)$ and we can, for instance, solve the equations
(\ref{perpendicular EM}) and (\ref{parallel EM}) near the boundary
$u\to0$ using Frobenius expansion
$E=u^\lambda\sum_{n=0}^\infty{a_nu^n}$ where the indicial equation has
solutions for $\lam=0,1$.

To solve the equations (\ref{perpendicular EM2}),~(\ref{parallel EM})
explicitly shall be the main part of this note. The solutions to these 5-d AdS equations of motion then give the 4-d field theory two point correlation as the functional derivative with respect to the boundary values of the on-shell AdS action
\begin{equation}
	\left.S=-\frac{1}{4g_B^2}\int{d^4xdu~\sqrt{g}e^{-cu}F_{AB}F^{AB}}\right|_{\mbox{on-shell}},
\end{equation}
with $g_B^2=16\pi^2R/N_c^2$. Considering $V_u=0$ gauge, we can write this as
\begin{eqnarray}
	S_{\mbox{on-shell}}&=&-\frac{N_c^2}{32\pi^2R}\left.\int_{-\infty}^\infty d^4x\left(\sqrt{g}e^{-cu}V_\mu F^{u\mu}\right)\right|_{u=0}^{u=1} \non
&=&\frac{N_c^2T^2}{16}\left.\int_{-\infty}^\infty d^4x~e^{-cu}\left(V_t\partial_uV_t-fV_i\partial_uV_i\right)\right|_{u=0}^{u=1}.
\end{eqnarray}
Fourier transforming to momentum space and selecting the particular direction chosen previously, we can rewrite the action using Minkowskian prescription formulated by Son and Starinets \cite{son02}. Together with the boundary condition that the solution of equations (\ref{perpendicular EM}) and (\ref{parallel EM}) must satisfy the incoming-wave boundary condition at the horizon $u=1$, the resulting on-shell action becomes
\begin{eqnarray}
\label{boundary action}
 S_{\mbox{on-shell}}&=&\frac{N_c^2T^2}{16}\lim_{u\to 0}\int\frac{d\omega~ dq}{(2\pi)^2}e^{-cu}\nonumber\\
&&\left[\frac{f}{q^2f-\omega^2}\partial_uE_\parallel(u,K)E_\parallel(u,-K)-\frac{f}{\omega^2}\partial_uE_\perp(u,K)E_\perp(u,-K)\right].\nonumber\\
\end{eqnarray}
From Eq. (\ref{boundary action}) and the condition described above, we
can now compute the retarded current-current correlation function in
term of two independent scalar functions\footnote{For a more detailed derivation of these functions see \cite{kovtun05}.}
\begin{eqnarray}
\label{longitudinal retarded Green function}
 \Pi^L(K)&=&-\frac{N_c^2T^2}{8}\lim_{u\to 0}\frac{\partial_uE_\parallel(u,K)}{E_\parallel(u,K)}, \\
\label{transverse retarded Green function}
 \Pi^T(K)&=&-\frac{N_c^2T^2}{8}\lim_{u\to 0}\frac{\partial_uE_\perp(u,K)}{E_\perp(u,K)}.
\end{eqnarray}
These functions in turn give us the photon and dilepton production at strong coupling via eq.~(\ref{eq:15}) and eqs. (\ref{eq:2}) and (\ref{eq:1}).

\section{Solving the system}
In this section we will solve the equations (\ref{perpendicular EM2})
and (\ref{parallel EM}) in order to compute the two scalar functions
(\ref{longitudinal retarded Green function}) and (\ref{transverse
  retarded Green function}). Furthermore, we will take the imaginary
part of those scalar functions and obtain the spectral density
function (\ref{eq:22}) for finite temperature system. 

The solutions which satisfy the incoming-wave boundary condition can be written in general as a Frobenius expansion near $u\to1$ 
\begin{equation}
\label{general solution}
 E_i(u)=(1-u)^{-i\omega/2}y_i(u),
\end{equation}
with $y_i(u)$ regular at $u=1$. We will solve and discuss these equations extensively for lightlike momenta relevant for photon-production, both semi-analytically for asymptotically small and large frequency and numerically for various values of the cut-off $c$ for the full range of momenta. For timelike and spacelike momenta we only present the numerical solution. 

\subsection{Lightlike momenta}
As has been explained in section 2, the longitudinal part of the scalar functions vanishes for lightlike momenta and we just need to compute the transverse part. 

\subsubsection{Analytic solutions for lightlike momenta at low and high frequency}

We are mainly interested in the effect of the IR-cut-off on photon production as compared to the previous AdS photon production calculation for scale-invariant $\cN=4$ SYM \cite{huot06}. In the low-frequency limit where its effect should be largest, we can solve (\ref{perpendicular EM}) perturbatively using $\omega\ll1$ as a small parameter. As noted in \cite{huot06}, there is a shortcut to do so. Given the two independent solutions $\phi_1 \pm i \phi_2$ to the differential equation $\phi''+ A(x)\phi'+B(x)\phi=0$, the Wronskian times $\exp(\int^x A(x'))$ is strictly conserved
\begin{eqnarray}
  \label{eq:5}
  \pa_x\left( e^{\int^x A(x')}\left[\bar{\phi}\pa_x\phi-\phi\pa_x\bar{\phi}\right]\right)=0~.
\end{eqnarray}
The transverse scalar can be rewritten as 
\begin{eqnarray}
\label{transverse scalar}
 \Pi^T(K)&=&\lim_{u\rar 0} \Pi^T(u,K) ~,\non
 \Pi^T(u,K)&\equiv& -\frac{N_c^2T^2}{8}\left[e^{-cu}(1-u^2)\frac{\bar{E}_{\perp}(u,K)}{\bar{E}_{\perp}(0,K)}\pa_u\frac{E_{\perp}(u,K)}{E_{\perp}(0,K)}\right].
\end{eqnarray}
The imaginary part of the transverse scalar $\Pi^T(u,K)$ is then
propotional to the conserved Wronskian and therefore independent of
the radial coordinate $u$:
\begin{equation}
\partial_u\Im[\Pi^T(u,K)]=0~. 
\end{equation}
With this fact, we can evaluate the imaginary part of (\ref{transverse scalar}) at any given value of $u$ which is convenient to our calculation. Let us choose $u=1$. Because the transverse scalar (\ref{transverse scalar}) contains an explicit factor of $(1-u)$, only the pole in $\bar{E}_{\perp}\pa_u E_{\perp}$ will contribute. Recalling that for any finite frequency $\ome$ the boundary conditions determine $E_{\perp}(u)$ to be of the form (\ref{general solution}), we immediately see that the undetermined regular part $y$ contains no pole by definition. Therefore without needing to solve the equation motion we see that
\begin{eqnarray}
  \label{eq:6}
  \Pi^T(1,K) = \frac{-N_c^2T^2}{8} \left(\frac{-i\ome}{2}\right) \left[ 2 e^{-c}\left|\frac{y(1)}{y(0)}\right|^2 \right]~.
\end{eqnarray}
The leading term in the limit $\ome \ll 1$ is the $\omega$-independent contribution to $|y(1)/y(0)|$. The determining equation (\ref{perpendicular EM2}) simplifies in that limit to effectively the first order equation (recall that $\ome=q$ for lightlike momenta)
\begin{eqnarray}
  \label{eq:7}
  \pa_u \pa_u E_{\perp} + (\pa_u (\ln f -cu))\pa_u E = 0 + \cO(\ome^2)~.
\end{eqnarray}
The incoming wave boundary condition demands that the $\ome=0$ solution be regular at $u=1$. Since $f =(1-u)(1+u)$, this solution is the trivial constant one. Therefore
\begin{eqnarray}
  \label{eq:8}
  \Pi^T(1,K) = \frac{i\ome N_c^2T^2}{8}e^{-c} + \cO(\ome^2)~.
\end{eqnarray}
In Appendix \ref{sec:spectr-funct-low} we compute the same answer directly by solving the differential equation perturbatively in $\ome$, which shows explicitly that $E_{\perp}(u)= \mbox{constant} + \cO(\ome)$ is indeed the correct solution to the boundary conditions.

Given $\Pi(1,K)$, the trace of spectral density function at low-frequency limit for lightlike momenta in photon production is proportional to its $u$-independent imaginary part
\begin{eqnarray}
\label{low-frequency spectral density for light-like}
 \chi^\mu_\mu(\omega=q)&=&-4\mbox{~Im}(\Pi^T(\omega=q))\nonumber \\
&=&\frac{\omega N_c^2T^2}{2}e^{-c}+\cO(\omega^2).
\end{eqnarray}
For $c=0$, we reproduce back the result from~\cite{huot06} at the first order. The vanishing of $c$ corresponds to either the limit $T\to\infty$ or to removing the IR scale $\Lambda_{IR}$. We see explicitly our intuition confirmed that the trace of spectral density at low-frequency depends on the cutoff parameter $c$, while simultaneously reproducing the $\cN =4$ result at high $T$. 

\bigskip 
At high-frequencies we do not expect the IR-cut-off to have a major effect. Let us show that to leading order the spectral function is in fact independent of the value of $c$ as one would expect. In this limit $\omega\gg1$, the argument leading up to eq. (\ref{eq:6}) does not hold\footnote{Note e.g. that in the singular term $(1-u)^{-i\ome/2}$ the order of limits $u\rar 1$ and $\ome \rar \infty$ do not commute.} and one cannot obtain the answer without solving the equation of motion (\ref{perpendicular EM2}).  Following \cite{huot06}, we will use the Langer-Olver method~\cite{olver54,olver97} to find the solution. The first step is to redefine 
\begin{equation}
 E_\perp(u)=\frac{e^{cu/2}}{\sqrt{-f(u)}}y(u)
\end{equation}
 for equation (\ref{perpendicular EM2}) and rewrite it as
\begin{equation}
\label{high-frequency eqm}
 y''(x)=[\omega^2 H(x)+G(x)]y(x),
\end{equation}
where $H(x)=\frac{x}{f(x)^2}$ and $G(x)=\frac{c^2}{4}-\frac{cx}{f(x)}-\frac{1}{f(x)^2}$ with $x=-u\in[-1,0]$.
For large $\omega$ the first term on the RHS dominates. Since it has a simple zero at $x=0$, we can transform Eq. (\ref{high-frequency eqm}) to Airy's equation plus terms subleading in $\ome$.
To do so, we introduce a new independent variable $\zeta$ and change variables to
\begin{eqnarray}
 \zeta\left(\frac{d\zeta}{dx}\right)^2&=& H(x) = \frac{x}{(1-x^2)^2}~.
\end{eqnarray}
Choosing conditions $\zeta(0)=0$ and $\zeta'(0)>0$ determines $\zet$ to be
\begin{equation}
 \zeta=\left[\frac{3}{2}\int_0^x\sqrt{H(t)}dt\right]^{2/3}.
\end{equation}
Rescaling $y(x)$ to
\begin{eqnarray}
 y&=&\left(\frac{d\zeta}{dx}\right)^{-1/2}W~,
\end{eqnarray}
eq. (\ref{high-frequency eqm}) becomes
\begin{equation}
\frac{d^2W}{d\zeta^2}=[\omega^2\zeta+\psi(\zeta)]W,
\end{equation}
with
\begin{equation}
 \psi(\zeta)=\frac{5}{16\zeta^2}+\frac{\left[4H(x)H''(x)-5H'^2(x)\right]}{16H^3(x)}\zeta+\frac{\zeta G(x)}{H(x)}
\end{equation}
For large $\ome$ we may ignore $\psi(\zet)$ and the equation reduces to Airy's equation. To leading order the solution is thus
\begin{equation}
  W(\zeta)=A_0\mbox{Ai}(\omega^{2/3}\zeta)+B_0\mbox{Bi} 
(\ome^{2/3}\zet)+\ldots,
\end{equation}
The incoming-wave boundary conditions at the horizon imply that $B_0$ should vanish. 
Thus the solution for $E_\perp(u)$ in asymptotic expansion for large $\ome$ is
\begin{equation}
 E_\perp(u) =
 \frac{A_0e^{cu/2}}{\sqrt{-f(u)}}\left[\frac{-u}{f(u)^2\zeta(-u)}\right]^{-1/4}\mbox{Ai}(\omega^{2/3}\zeta(-u))+\ldots ,
\end{equation}
and the transverse scalar at high-frequency limit equals
\begin{equation}
\label{high frequency correlator}
 \Pi^T = -\frac{N_c^2T^2}{8}\lim_{u\to0}\left(\frac{c}{2}+\frac{1}{4}\pa_u\ln\left(\frac{-\zet(-u)}{u}\right)+\frac{\pa_u\mbox{Ai}(\omega^{2/3}\zeta(-u))}{\mbox{Ai}(\omega^{2/3}\zeta(-u))}\right) +\ldots.
\end{equation}
Before we move on, it is helpful to expand $\zeta(-u)$ around $u=0$
\begin{equation}
\zeta(-u)=-(-1)^{2/3}u-\frac{2}{7}(-1)^{2/3}u^3+\cO(u^5)~.
\end{equation}
Therefore the middle term in (\ref{high frequency correlator}),
\begin{eqnarray}
  \label{eq:9}
  \pa_u\ln\left(\frac{-\zet(-u)}{u}\right) = \frac{1}{(-1)^{2/3}+..} \left(\frac{6}{7}(-1)^{2/3} u +\ldots\right),
\end{eqnarray}
vanishes as $u \to 0$. Knowing the asymptotics of the Airy function
 the last term of (\ref{high frequency correlator}) can be written as
\begin{eqnarray}
 \lim_{u\to0}\frac{\mbox{Ai}'(\omega^{2/3}\zeta(-u)))}{\mbox{Ai}(\omega^{2/3}\zeta(-u)))}&=&-(-\omega)^{2/3}\frac{\mbox{Ai}'(0)}{\mbox{Ai}(0)}\nonumber\\
&=&(-\omega)^{2/3}\frac{3^{1/3}\Gamma(2/3)}{\Gamma(1/3)},
\end{eqnarray}
and thus we obtain 
\begin{equation}
 \Pi^T = -\frac{N_c^2T^2}{8}\left(\frac{c}{2}+\frac{e^{2\pi i/3}\omega^{2/3}3^{1/3}\Gamma(2/3)}{\Gamma(1/3)}\right).
\end{equation}
Note that this transverse scalar therefore depends on $c$. However, only the real part does. The trace of the spectral density function in high-frequency limit for lightlike momenta 
\begin{eqnarray}
\label{spectral at high-frequency}
 \chi^\mu_\mu&=&-4~\mbox{Im}(\Pi^T) \nonumber\\
&\sim&\frac{N_c^2T^2}{4}\frac{\omega^{2/3}3^{5/6}\Gamma(2/3)}{\Gamma(1/3)}.
\end{eqnarray}
does not depend on the cutoff parameter $c$ at least up to first order
and yields the same result as the calculation in $\cN=4$ SYM. The fact
that $c$ does appear in the real part of the transverse scalar
indicates that at first subleading order the spectral density function
will likely differ from the $\cN=4$ result. The numerical
  results in the next section bear this out.

\subsubsection{Numerical solution for lightlike momenta}

The analytic asymptotic solutions are a guidance to the full spectral
function. The full solutions of equation (\ref{perpendicular EM}) for
non-zero $c$ are very difficult to find, as we remarked earlier. This is due to the irregular singular point at $u=\infty$ for $c\neq 0$ where  analytic solutions are not known. In this subsection we are going to look for numerical solutions for non-zero $c$. 

We start from the general solution (\ref{general solution}) which satisfies 
the incoming wave boundary condition. To set a parametrization of the
initial conditions for the $u=1$ regular function $y_i(u) =
E_i(1-u)^{i\ome/2}$ of Eq. (\ref{general solution}), we write the general solution as a polynomial expansion around $u=1$, $y(u)=\sum_{n=0}^\infty a_n(1-u)^n$. Substituting (\ref{general solution}) into equation (\ref{perpendicular EM2}) for lightlike momenta, we obtain the equation
\begin{eqnarray}
 &&\sum_{n=0}^{\infty} \left[a_n\left(n-i\frac{\omega}{2}\right)^2(1-u)^{n-2}+c~a_n\left(n-i\frac{\omega}{2}\right)(1-u)^{n-1}\right.\nonumber \\
&&  \left.-\sum_{m=0}^{\infty}\left[\frac{a_n}{2^{m+1}}\left(n-i\frac{\omega}{2} +\frac{\omega^2(m+1)}{4}\right)(1-u)^{n+m-1}-\frac{a_n\omega^2}{2^{m+2}}(1-u)^{n+m-2}\right]\right]=0.
\end{eqnarray}
The second sum (over $m$) arises from expanding $\frac{1}{1+u}=\sum_{n=0}^\infty\frac{1}{2^{n+1}}(1-u)^n$ and $\frac{1}{(1+u)^2}=\sum_{n=0}^\infty\frac{(n+1)}{2^{n+2}}(1-u)^n$. In order to find the coefficients $a_n$, we have to solve this equation for each power of $(1-u)$ and obtain
\begin{eqnarray}
 (1-u)^{-2}&:&~~a_0~\mbox{(arbitrary)}, \nonumber \\
 (1-u)^{-1}&:&~~a_1=\frac{i\omega(c-1/2)}{2(1-i\omega)}a_0, \nonumber \\
\vdots&&\nonumber \\
 (1-u)^{k-2}&:&~~a_k=f_k(\omega,c)a_0,
\end{eqnarray}
with $f_k$ are functions of $\omega$ and $c$ which vanish at $\omega=0$. This gives us $y(u)$ and $y'(u)$ at $u=1$ in terms of the above coefficients
\begin{eqnarray}
\label{bc for p at u=1}
 y(1)&=&a_0,\nonumber \\
 y'(1)&=&-a_1=-a_0\frac{i\omega(c-1/2)}{2(1-i\omega)}.
\end{eqnarray}
These will be the two initial conditions for the differential equation for $y(u)$. The explicit differential equation it must satisfy is
\begin{eqnarray}
\label{eq:10}
  &&y''+\left(\frac{i\omega}{1-u}-\frac{2u}{1-u^2}-c\right)y' \non
&&+\left[\frac{\omega^2u}{(1-u^2)^2}+\frac{2i\omega-\omega^2}{4(1-u)^2}-\frac{i\omega}{2(1-u)}\left(\frac{2u}{1-u^2}+c\right)\right]y=0.
\end{eqnarray}
Notice that the initial conditions for $y(u)$ still depend on an arbitrary constant $a_0$. Physical quantities, such as the spectral density function, depend on ratios of $y(u)$ and its derivatives and are independent of this constant. We are therefore free to set it to any value; we will choose $a_0=1$.

Let us express the trace of spectral density function in terms of $y(u)$:
\begin{equation}
 \chi_\mu^\mu=\frac{N_c^2T^2}{2}\left(\frac{\omega}{2}+\mbox{Im}\left(\frac{y'(0)}{y(0)}\right)\right).
\end{equation}
Alternately we could use the modified Wronskian formulation for $\Pi^T((K)$, eq. (\ref{transverse scalar}) and evaluate it at $u=1$. An equivalent expression for the trace of spectral density function in this limit becomes
\begin{eqnarray}
 \chi_\mu^\mu
&=&\frac{\omega N_c^2T^2}{2}e^{-c}\frac{|y(1)|^2}{|y(0)|^2}.
\end{eqnarray}

\subsubsection{The spectral density for lightlike momenta}
\label{sec:spectr-dens-lightl}

\begin{figure}[htp]
  \begin{center}
    \includegraphics[width=12cm]{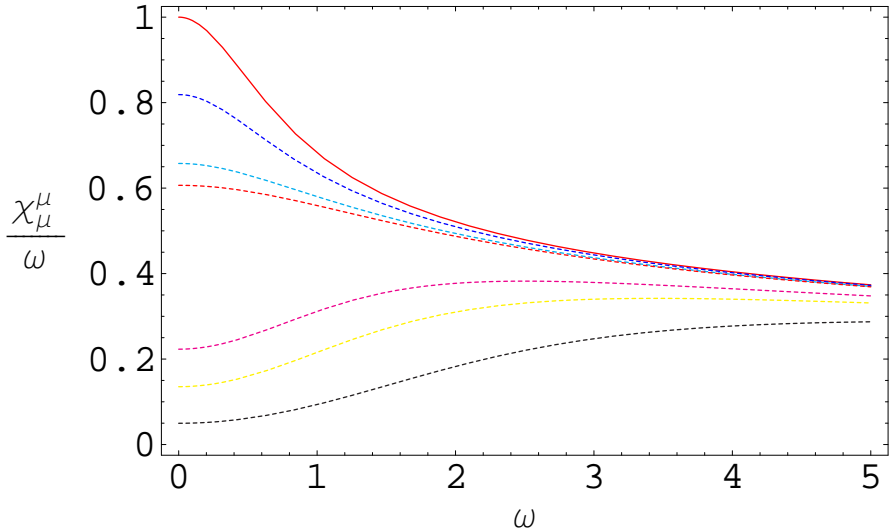}
  \end{center}
  \caption{\small Trace of the spectral function for lightlike momenta in units of $\frac{1}{2}N_c^2T^2$, plotted as a function of frequency with $\omega\equiv k^0/(2\pi T)$. The solid line (red) shows the exact result for $c=0$ and the dashed lines downward show numerical analysis for $c=0.2, 0.419035, 0.5, 0.6, 1.5, 2, 3$.}  
\label{graph light-like spectral/w}
\end{figure}

Solving eq. (\ref{eq:10}) numerically with initial conditions (\ref{bc for p at u=1}), we find the spectral density function $\chi_\mu^\mu$ for lightlike momenta for various values of the IR-cut-off $c$.\footnote{Numerical solutions were
  obtained using the NDSolve routine in Mathematica.}  The results are shown in Fig. \ref{graph light-like spectral/w} and we clearly see the dependency at low frequencies on the IR-cut-off. The behaviour at high-frequency on the other hand appears less and less sensitive.
What is remarkable is the similarity between this soft-wall AdS/QCD
result, Fig \ref{graph light-like spectral/w}, for
the trace of the spectral function for light-like momenta and
 of Mateos and Pati\~{n}o for massive flavor deformations of
the AdS dual of $\cN=2$ theories, Fig. 3 in \cite{mateos07}.
As we will discuss in section \ref{sec:concl-soft-wall}, this
similarity can be explained by relating the two
computations. Inherently this then
partially validates the soft-wall AdS/QCD model.

There is, however, one fundamental difference between the result here and the massive $\cN=2$ computation. Both models are thermodynamically unstable for large IR-cut-off, signalling the transition back to the confining regime. In the $\cN=2$ model this is clearly illustrated by the appearance of thermal resonances in the spectral function when formally evaluated beyond the critical cut-off. Fig1. shows that in AdS/QCD these resonances remain absent beyond the critical value $c>0.419035$ \cite{herzog06}. The absence of thermal resonances was presaged by Huot et al. \cite{huot06}. Realizing that their results for photon production in the AdS dual of pure $\cN=4$ SYM are unaffected by a hard-wall IR-cut-off, they speculated that this would be generic. It was premised on the fact that in the hard-wall case, the IR-cut-off is always inside the horizon. Rough dimensional analysis illustrates that the soft-wall case is similar: at the transition the cut-off scale $c^{-1} \simeq 2.5 $ is beyond the horizon $u=1$. However, a similar argument holds for the massive $\cN=2$ AdS dual. As we discuss in section 4, the real reason for the absence of thermal resonances is probably simply that an blunt soft- or hard- IR-cut-off is too crude to capture this information.

\subsection{Timelike and spacelike momenta}
For time and space-like momenta, both $\Pi^T$ and $\Pi^L$ can contribute to spectral density function $\chi_\mu^\mu(K)$. Also for these cases, the mode equations (\ref{perpendicular EM}) and (\ref{parallel EM}) cannot be solved analytically for arbitrary frequency($\omega$) and wave vector($q$), and we determine the spectral function numerically.

\subsubsection{Numerical solution for transverse scalar function}

Following the same procedure in numerical analysis for lightlike momenta above, we substitute the general solution (\ref{general solution}) for transverse direction into (\ref{perpendicular EM}) and obtain an equation for $y(u)$
\begin{eqnarray}
\label{eq:11}
 &&y_\perp''+\left(\frac{i\omega}{1-u}-\frac{2u}{1-u^2}-c\right)y_\perp'\non
&&+\left[\frac{\omega^2-q^2(1-u^2)}{u(1-u^2)^2}+\frac{2i\omega-\omega^2}{4(1-u)^2}-\frac{i\omega}{2(1-u)}\left(\frac{2u}{1-u^2}+c\right)\right]y_\perp=0.
\end{eqnarray}
As in the lightlike case, to determine the initial conditions we expand $y(u)=\sum_{n=0}^{\infty}a_n(1-u)^n$ around $u=1$, with 
\begin{eqnarray}
\label{eq:13}
a_0~\mbox{(arbitrary)}, ~~a_1=\frac{\omega^2-q^2+i\omega(1/2-c)}{2(i\omega-1)}a_0~,~~a_k=f_k(\omega,q,c)a_0,
\end{eqnarray}
where again $f_k$ are functions of $\omega,q$ and $c$ which vanish at $\omega=q=0$. Using the modified Wronskian extension the imaginary part of transverse scalar function is therefore given by
\begin{equation}
 \mbox{Im}(\Pi^T(K))=-\frac{\omega N_c^2T^2}{8}e^{-c}\frac{|y_\perp(1)|^2}{|y_\perp(0)|^2},
\end{equation}
with $y(u)$ a solution to eq. (\ref{eq:11}) with initial conditions determined from Eq.~(\ref{eq:13}).

\subsubsection{Numerical solution for longitudinal scalar function}

Substitute (\ref{general solution}) into the equation of motion for the longitudinal direction (\ref{parallel EM}), we obtain
\begin{eqnarray}
\label{eq:12}
 &&y_\parallel''+\left(\frac{i\omega}{1-u}-\frac{2u\omega^2}{(1-u^2)(\omega^2-q^2(1-u^2))}-c\right)y_\parallel'+\left[\frac{\omega^2-q^2(1-u^2)}{u(1-u^2)^2}\right.\nonumber\\
&&+\left.\frac{2i\omega-\omega^2}{4(1-u)^2}-\frac{i\omega}{2(1-u)}\left(\frac{2u\omega^2}{(1-u^2)(\omega^2-q^2(1-u^2))}+c\right)\right]y_\parallel=0.
\end{eqnarray}
Expanding $y_\parallel(u)= \sum_{n=0}^{\infty}a_n(1-u)^n$ around $u=1$ gives us 
\begin{eqnarray}
\label{eq:14}
a_0~\mbox{(arbitrary)}~,~~a_1=\frac{\omega^2-q^2+i\omega\left(\frac{1}{2}-c-\frac{2q^2}{\omega^2}\right)}{2(i\omega-1)}a_0~,~~a_k=f_k(\omega,q,c)a_0,
\end{eqnarray}
where again $f_k$ are functions of $\omega,q$ and $c$ which vanish at $\omega=q=0$. The imaginary part of the longitudinal scalar function is
\begin{equation}
 \mbox{Im}(\Pi^L(K))=-\frac{\omega N_c^2T^2}{8}\left(\frac{1}{2}+\mbox{Im}\left(\frac{y_\parallel'(0)}{\omega y_\parallel(0)}\right)\right),
\end{equation}
with $y_{\parallel}(u)$ the solution to (\ref{eq:12}) with initial conditions determined from Eq.~(\ref{eq:14}).

\bigskip
\subsubsection{The spectral density for time- and space-like momenta}
Following formula (\ref{eq:15}), we can now write the trace of spectral function for time- and space-like momenta as
\begin{eqnarray}
 \chi_\mu^\mu(K)=\frac{\omega N_c^2T^2}{2}\left[e^{-c}\frac{|y_\perp(1)|^2}{|y_\perp(0)|^2}+\frac{1}{4}+\frac{1}{2\omega}\mbox{Im}\left(\frac{y_\parallel'(0)}{y_\parallel(0)}\right)\right].
\end{eqnarray}
The complete results for $\chi_\mu^\mu$ are plotted in Fig.\ref{graph
  time- and space-like spectral/w} and Fig.\ref{graph time- and
  space-like spectral/w1} as a function of frequency for several
values of the spatial momentum.
As we increase the value for $c$, one clearly sees that at low momenta the function decreases compared to $c=0$. 

\begin{figure}[ht]
\begin{minipage}[t]{0.5\linewidth}
\centering
    \includegraphics[width=7cm]{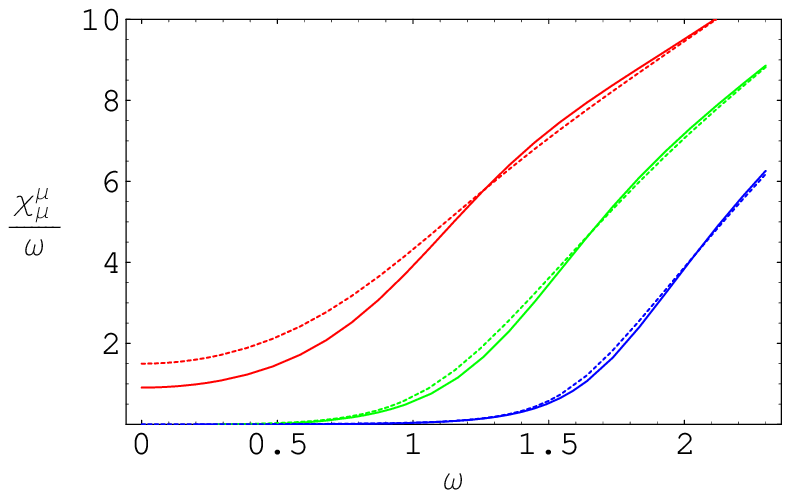}
\caption{\small Spectral function trace $\chi_\mu^\mu/\omega$, in units of $N_c^2T^2/2$, plotted as a function of $\omega$. The solid lines describe $c=0.5$ and the dashed lines for $c=0$ while different colors represent $q=0$(red), $q=1$(green), and $q=1.5$(blue).}  
\label{graph time- and space-like spectral/w}
\end{minipage}
\hspace{0.5cm}
\begin{minipage}[t]{0.5\linewidth}
\centering
    \includegraphics[width=7cm]{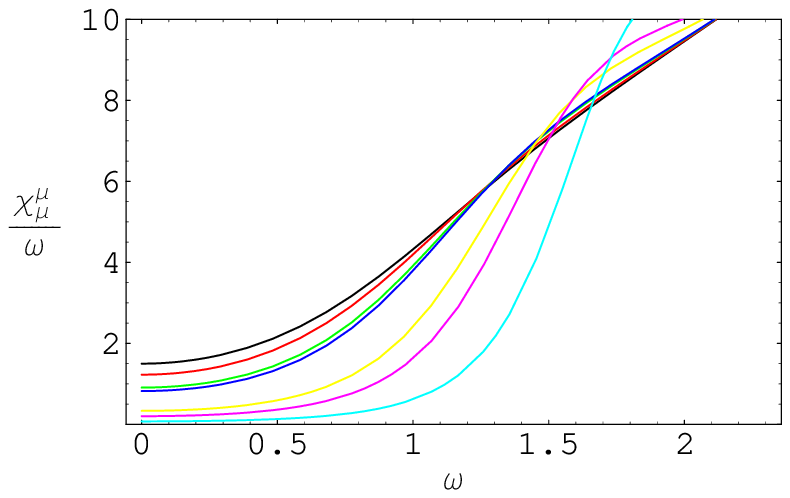}
\caption{\small Spectral function trace $\chi_\mu^\mu/\omega$, in units of $N_c^2T^2/2$, plotted as a function of $\omega$ for $q=0$ and various values of $c=0$\mbox{(black)}, $c=0.2$\mbox{(red)}, $c=0.5$\mbox{(green)}, $c=0.6$\mbox{(bue)}, $c=1.5$\mbox{(yellow)}, $c=2$\mbox{(magenta)}, and $c=3$\mbox{(cyan)}.}
\label{graph time- and space-like spectral/w1}
\end{minipage}
\end{figure}

\subsection{Electrical conductivity}

With the spectral density in hand, it is now straightforward to compute the electrical conductivity $\sigma$. Here, we will use Eq. (\ref{Kubo formula2}) as we have an analytic expression of the spectral density for lightlike momenta.
Substituting (\ref{low-frequency spectral density for light-like}) into  (\ref{Kubo formula2}) yields
\begin{eqnarray}
\label{electrical conductivity}
 \sigma&=&\lim_{k^0\to0}\frac{e^2}{4T}\frac{\chi^\mu_\mu(\omega=q)}{e^{k^0/T}-1}\nonumber\\
 &=&\lim_{k^0\to0}\frac{e^2}{8\pi}\frac{N_c^2\exp(-c)k^0(1+O(k^0))}{k^0/T(1+O(k^0))}\nonumber\\ 
 &=&e^2\frac{N_c^2T}{16\pi}\exp(-c),
\end{eqnarray}
with $e$ the electric charge. We again note the presence of the scaling factor $e^{-c}$ which dampens the IR-properties, including charge diffusion, of the system. Note in particular that this IR-suppresion is also present in the charge susceptibility $\Xi = N_c^2T^2c/8(e^c-1)$ and the more ``universal'' diffusion constant $D\equiv \sigma/e^2\Xi = (1-e^{-c})/2\pi T c$ (see Appendix \ref{sec:comp-diff-const}). Physically this makes sense, as a mass-gap should dampen any hydrodynamic behaviour and the general AdS/CFT computation for scale-dependent currents
  \begin{eqnarray}
    \label{eq:20}
	\left.S_{AdS} \sim \int{d^4xdu~\sqrt{-g}\frac{1}{g^2_{\rm eff}(u)}F_{AB}F^{AB}} \right.,  
  \end{eqnarray} 
demonstrates this explicitly \cite{Kovtun:2003wp}\footnote{This suggests a trivial violation of the PSS shear-viscosity-bound by IR-suppressing hydrodynamic behaviour. As the derivation of the viscosity in \cite{Kovtun:2003wp} suggests, however,  and the explicit computation in massive $\cN=2$ models shows \cite{Mateos:2006yd}, this is not case.}
\begin{eqnarray}
  \label{eq:21}
  D \sim \frac{1}{g^{2}_{\rm eff}(u=1)}\int_0^1 du\, g^{2}_{\rm eff}(u)\ldots 
\end{eqnarray}

\section{Conclusion: Soft wall cut-offs as an IR mass-gap.}
\label{sec:concl-soft-wall}

The essential new ingredient in Soft-wall AdS/QCD is the ad-hoc
cut-off of the radial AdS-direction. It is intended to capture the
dominant effects of the scale dependence of QCD \cite{erlich05,herzog06}. 
However, its ad-hoc introduction opens it to criticism; especially when interpreted as a dilaton-profile without taking into account back-reaction effects or the dilaton equation of motion (see the footnote in the introduction). On the other
hand the succesful results of the model~\cite{karch06,herzog06}, suggest that it does capture
the essential IR behaviour correctly. 

The result for AdS/QCD photon production supports this further. As previously emphasized it closely resembles photon production due to quarks 
for $\cN=2$ theories with massive
flavor in the probe approximation $N_f \ll N_c$ \cite{mateos07}. These 
theories
descend from brane-constructions in string theory, and therefore
have no ad hoc component to criticise. Recall that in these theories, the probe
approximation means that one may consider the
flavor group as a global symmetry. The $U(1)$ theory with respect to which
photons are defined is a subgroup of this group and the tunable quark mass --- a free parameter in the brane construction --- functions as the scale in these theories.
On the other hand, because the
matter and symmetry content is different from QCD, one could question
how relevant massive $\cN=2$ SQCD results are to reality. The observation
we make now is that the resemblence between the trace of the spectral function
$\chi^{\mu}_{\mu}$ in these $\cN=2$ SQCD theories as a function of the
quark mass $m$ and the AdS/QCD spectral function as a function of the
IR-cut-off $c$ can be mathematically explained. Both therefore
demonstrate again that AdS/CFT results are remarkable universal and
robust across fundamentally different theories. This is therefore
strong support for soft-wall AdS/QCD, despite its ad-hoc
IR-cut-off, as well as massive $\cN=2$ SQCD, despite its unrealistic
matter content, as descriptions of QCD.

To relate the massive $\cN=2$ SQCD result to AdS/QCD, we note that Mateos and Pati\~{n}o showed that in
$\cN=2$ SQCD 
the defining equation relevant for the trace of the spectral function
for lightlike momenta can be
deduced from an action\footnote{We only consider the D3/D7 brane set-up of \cite{mateos07}. The gauge/gravity duality for the D4/D6 brane set-up they also consider is not yet fully understood.} 
\begin{eqnarray}
  \label{eq:1a}
  S \sim \int dudx_0dx_1 \left[-P(u)(\pa_0
  V_{\perp})^2+fP(u)(\pa_1V_{\perp})^2+Q(u)(\pa_u V_{\perp})^2\right]~,
\end{eqnarray}
where 
\begin{eqnarray}
  \label{eq:2a}
  P(u) &=& \frac{u^3\sqrt{g(\psi_{m,0}(u),u)}}{uf}~,\non
~~Q(u) &=& f \frac{(1-\psi_{m,0}^2(u))^3}{u^3\sqrt{g(\psi_{m,0}(u),u)}}\non 
       &=&\frac{1}{u} \frac{u^3f\sqrt{g(\psi_{m,0}(u),u)}}{uf}\frac{u^2f^2(1-\psi_{m,0}^2(u))^3}{u^6g(\psi_{m,0}(u),u)}~.
\end{eqnarray}
Here $f=f(u)=(1-u^2)$ is the non-extremality function in the D3-brane metric (\ref{AdS black hole}). The function $\psi_{m,0}(u)$ is the solution to the embedding equation of motion for the D7-flavor brane derived from the DBI-action
\begin{eqnarray}
  \label{eq:3}
  S \sim \int du \sqrt{g(\psi_m(u),u)} =\int du \frac{1}{u^3}(1-\psi_m^2)\sqrt{1-\psi^2+4u^2f\psi^{'2}}~,
\end{eqnarray}
i.e. $g(\psi(u),u)$ is the induced metric on the flavor brane. The
$u=0$ boundary behavior of the solution $\psi_{m,0} = \frac{m}{\sqrt{2}} u^{1/2} + \Lam
u^{3/2}+\ldots$ is determined by the
masses $m$ and condensate expectation value $\langle qq\rangle \sim
\Lam$ of the quarks. For the massless theory $\psi_{m=0,0}=0$ and $\sqrt{g} =u^{-3}$. Thus to find
the spectral function, one must first solve the differential equation
for $\psi_m(u)$ with the appropriate boundary conditions and then solve the
differential equation for $V_{\perp}$ \cite{mateos07}. The first step correctly
incorporates the backreaction of the modified IR-physics as opposed to the AdS/QCD ad-hoc cut-off.

The massive case $\psi_{m,0}(u)\neq0$ is therefore a step more
involved than the massless case, unlike AdS/QCD where the scale is a
mild modification $c\neq 0$ of the defining differential equation
(\ref{perpendicular EM2}). However, searching for a closer match, one
quickly realizes that the massless equation
(for lightlike momenta $\ome=\vec{k}$),
\begin{eqnarray}
  \label{eq:16}
  \pa_u^2V_{\perp}+\pa_u(\ln Q)\pa_uV_{\perp}+ \vec{k}^2(1-f)\frac{P}{Q}V_{\perp}  &=&0 \non
\Rightarrow \pa_u^2 V_{\perp} + \pa_u(\ln\left(f\right))\pa_uV_{\perp} + \vec{k}^2(1-f)\frac{(uf)^{-1}}{f} V_{\perp} &=& 0~,
\end{eqnarray}
is exactly the AdS/QCD equation (\ref{perpendicular EM2}) for $c=0$
and we are therefore lead to consider a change of variables for the massive case that resembles that of the massless case. 
\def\tu{\tilde{u}}
Thus we define a new variable $\tu$ such that
\begin{eqnarray}
  \label{eq:17}
du \frac{u^3  \sqrt{g(\psi_{m,0}(u),u))}}{uf} = d\tu \frac{1}{\tu\tilde{f}} 
\end{eqnarray}
with $\tilde{f}\equiv f(\tilde{u})$. By construction the parameter $P$
in the new variable 
is identical to the massless case and $Q$ is seen to be a mild modification
\begin{eqnarray}
  \label{eq:18}
  P(\tu) &=& \frac{1}{\tu f(\tu)}~, \non
  Q(\tu) &=& {f(\tilde{u})} \frac{\tu (1-\psi^2_{m,0})^3}{u(\tu)}~. 
\end{eqnarray}
Note that the solution to the massive embedding equation of motion,
$\psi_{m,0}\neq0$, is implicit in the transformation (\ref{eq:17}). In
this new variable, however, we
see, that its specific form only mildly modifies the massless
differential equation
\begin{eqnarray}
  \label{eq:19}
  \pa^2_{\tu} V_{\perp} + \pa_{\tu}\left[\ln(\tilde{f}) +\ln((1-\psi^2_{m,0})^3\frac{\tu}{u(\tu)})\right]\pa_{\tu} V_{\perp} + \vec{k}^2(1-\tilde{f})\frac{(\tu\tilde{f})^{-1}}{f(1-\psi^2)^3}V_{\perp} = 0~.
\end{eqnarray}
and the close relation to AdS/QCD is now apparent. The resemblance of
the spectral functions is especially explained, if
we recall that it is primarily determined by the $u=0$ behaviour of
the solution (\ref{transverse retarded Green function}).\footnote{One should be careful in that the change of coordinates (\ref{eq:17}) in principle will also change the boundary conditions one must impose.} 
As we know
what the $u=0$ behaviour of the solution $\psi_{m,0}=
\frac{m}{\sqrt{2}}u^{1/2}+\ldots $ must be, Eq. (\ref{eq:17}) shows 
that asymptotically
$\tilde{u}=u+\frac{m^2}{4}u^2+\ldots$ and we can putatively identify
the mass $m$ with the IR-cut-off $c$:
\begin{eqnarray}
  \label{eq:23}
  -c\tu ~~& \simeq & ~~\ln(1-\psi^2_{m,0})^3\frac{\tu}{u} = \ln
  (1-\frac{m^2}{2}\tu+\ldots)^3-\ln(1-\frac{m^2}{4}\tu+\ldots)\non
~~&\simeq & ~~-\frac{5}{4}m^2\tu+\ldots
\end{eqnarray}
The map between AdS/QCD and $\cN=2$ SQCD
is not exact; clearly we should not have expected it to be. 
The latter shows thermal resonances in the spectral function
for masses $m>1.3092$ which is the value beyond which the AdS
black-hole solution becomes thermodynamically unstable
\cite{mateos07}. The AdS/QCD description is much cruder as is no
resonances show up even beyond the unstable regime $c>0.419035$. These thermal resonances are encoded in the subtleties of the embedding function $\psi_{m,0}(u)$ which carries more information than just the mass as an IR-cut-off. Precisely, the embedding function determines whether the flavor D7-brane is in ``Minkowski embedding'' or ``black hole embedding''  corresponding to the low $T$ confining or high $T$ deconfining phase~\cite{mateos07}. Clearly, the $\cN=2$ SQCD theory has a more detailed description at the physics. On the other hand, the results here do
show that in the stable phase the simple AdS/QCD model
describes the IR-consequences of a mass-gap remarkably well and the
above derivation explains mathematically why. This in itself lends
support to continue to study AdS/QCD as a good toy model for
real-world physics.

\acknowledgments
We are grateful to D. Mateos for correspondence. KS thanks and acknowledges the hospitality of the Galileo Galilei Institute in Firenze. This research was supported in part by  a VIDI Innovative Research Incentive Grant from the Netherlands Organisation 
for Scientific Research (NWO).

\appendix

\section{Spectral function low frequency limit for lightlike momenta}
\label{sec:spectr-funct-low}

Here we find an analytic expression for the low-frequency limit of the
transverse scalar and spectral density for lightlike momenta by
solving the differential equation for the $E_{\perp}(u)$
perturbatively, rather than using the Wronskian shortcut, explained
above eq. (\ref{eq:5}). 

We first extract the other regular singularity at $u=-1$, writing
\begin{equation}
\label{ansatz}
	E_\perp(u)=(1-u)^{-i\omega/2}(1+u)^{-\omega/2}Y(u),
\end{equation}
with $Y(u)$ regular at $u=1$ and substitute this into (\ref{perpendicular EM}).
Changing variables to $v=1/2(1-u)$, we obtain the differential equation
\begin{eqnarray}
\label{equation near u=1}
v(1-v)Y''+\left[(1-i\omega)-(2-i\omega-\omega-2c)v-2cv^2\right]Y' &&\non
-\left\{\frac{1}{2}\left[-\omega-i\omega+i\omega^2\right]-c[\omega v-i\omega+i\omega v]\right\}Y&=&0.
\end{eqnarray}
In the absence of the IR-cutoff, $c=0$, we recognize a hypergeometric equation with solution  \cite{huot06}
\begin{equation}
 Y(u)={}_2F_1\left(1-\frac{1}{2}(1+i)\omega,-\frac{1}{2}(1+i)\omega;1-i\omega;\frac{1}{2}(1-u)\right).
\end{equation}
As we noted earlier, the presence of $c$ changes the nature of the equation and no formal solution is known. On physical grounds we expect the effects of $c$ to dominate the low frequency part of the spectral function. Expanding $Y(u)$ as 
\begin{equation}
\label{y expansion}
 Y=Y_0+\omega Y_1+\omega^2 Y_2+\omega^3 Y_3+\cdots,
\end{equation}
we find to first order in $\ome$,
\begin{eqnarray}
\label{first expansion}
 \omega^0~~:&~~~~~~~~~~~~~~~~~~~~~~~~~v(1-v)Y''_0+[1-2v+2cv(1-v)]Y'_0&=0, \\
\label{second expansion}
 \omega^1~~:&~v(1-v)Y''_1+[v-i(1-v)]Y'_0+[1-2v+2cv(1-v)]Y'_1&\nonumber\\
&~~~~~~~~~~~~~~~~~~~~~~~~~~~~~~
+\left\{\frac{1}{2}(1+i)+c[v-i(1-v)]\right\}Y_0&=0.
\end{eqnarray}
These two equations have solutions
\begin{eqnarray}
 Y_0(v)&=&A+B\left[e^{-2c}\mbox{Ei}(2c-2cv)-\mbox{Ei}(-2cv)\right],\\ 
Y_1(v)&=&C+\frac{A}{2}\left[\ln(v-1)+i\ln v\right]+\non
&& \left[e^{-2c}\mbox{Ei}(2c-2cv)-\mbox{Ei}(-2cv)\right]\left[D+\frac{B}{2}\left[\ln(v-1)+i\ln v\right]\right], \nonumber 
\end{eqnarray}
with $A,B,C,D$ constants of integration and
$\mbox{Ei}(x)=-\int^\infty_{-x}\frac{e^{-t}}{t}dt$ the exponential
integral function. To determine the integration constants, recall that
by construction the solutions must be regular as $v\to 0$ ($u \rar
1$). 
Since the exponential integral $\mbox{Ei}(v)$ diverges at $v=0$, we
must set $B=0$. 
To determine regularity of $Y_1(v)$, recall that $\mbox{Ei}(x)$ can be written as
\begin{equation}
 \mbox{Ei}(-x)=\gamma+\ln x+\sum_{n=1}^\infty{\frac{(-1)^nx^n}{n!n}},~~~~\mbox{for}~~x>0,
\end{equation}
with $\gamma$ the Euler-Mascheroni constant. Since the variable $v\in [0,1/2]$, and $c>0$, regularity at $v=0$ demands $D=iA/2$.
For convenience, let us also redefine the constant $C= i\tilde{C}A/2$.  Substituting those constants into $Y_1$, we obtain the solution for $E_{\perp}$ in the low frequency limit
\begin{eqnarray}
E_{\perp}(u) &=& A(1-u)^{-i\ome/2}(1+u)^{-\ome/2}\left\{ 1+ i\frac{\ome}{2}\left[\tilde{C}+e^{-2c}\mbox{Ei}(c(1+u))-\mbox{Ei}(c(u-1))\right.\right.\non
&&\left.\left.-i\ln\left(\frac{u+1}{2}\right)+\ln \left(\frac{1-u}{2}\right)\right]+\cO(\ome^2)\right\}.
\end{eqnarray}
Using the definition of the exponential integral function, we straightforwardly obtain the leading low-frequency contribution to transverse scalar function
\begin{eqnarray}
  \label{eq:4}
  \Pi^T(\ome=q) &=& -\frac{N_c^2T^2}{8} \left[-i\frac{\ome}{2} -\frac{\ome}{2} + \frac{i\ome}{2}(ce^{-2c}\mbox{Ei}'(c)-c\mbox{Ei}'(-c) -i-1) +\cO(\ome^2)\right] \non
&=& \frac{i\ome N_c^2T^2}{16} \left[-2i - (e^{-c}+e^{-c}) +\cO(\ome^2)\right].
\end{eqnarray}
This is the exact answer. The imaginary part computed via the conserved Wronskian shortcut  (\ref{eq:8}) clearly agrees.

\section{The susceptibility and the diffusion constant}
\label{sec:comp-diff-const}

We follow the procedure to compute the diffusion constant 
described in \cite{policastro02}. 
Using the gauge $V_u=0$, we can rewrite equation (\ref{eom V_t}) as 
\begin{equation}
\label{Vz in terms of V_t}
 V_{\parallel}
=\frac{uf}{q\omega}V_t''-c\frac{uf}{q\omega}V_t'-\frac{q}{\omega}V_t.
\end{equation}
Substituting into equation (\ref{V_u}) we obtain a 
second order differential equation for $E=V_t'$
\begin{equation}
 E''+\left[\frac{(uf)'}{uf}-c\right]E'+\left[\frac{\omega^2-q^2f}{uf^2}-c\frac{(uf)'}{uf}\right]E=0.
\end{equation}
Imposing the same incoming-wave boundary condition as before and
extracting the singularity at the horizon $u=1$, we rewrite
$E=(1-u)^{-i\omega/2}y$, where $y$ is a regular function at the
horizon. The function $y$ must obey the equation
\begin{eqnarray}
&& y''+\left[\frac{i\omega}{1-u}+\frac{(uf)'}{uf}-c\right]y'+\non
&&+\left[\frac{i\omega(i\omega+2)}{4(1-u)^2}+
\frac{i\omega((uf)'-cuf)}{2uf(1-u)}+\frac{\omega^2-q^2f}{uf^2}-c\frac{(uf)'}{uf}\right]y=0.
\end{eqnarray}
For low frequency and momentum, we again solve the equation
pertubatively in $\omega$ and $q$ 
\begin{equation}
 y(u)=y_{00}+\omega y_{10}+q^2 y_{02}+\cdots.
\end{equation}
Up to first order in $\omega$ and $q^2$, we find the system of equations
\begin{eqnarray}
 \omega^0q^0&:& y_{00}''+\left[\frac{(uf)'}{uf}-c\right]y_{00}'-c\frac{(uf)'}{uf}y_{00}=0, \nonumber \\
 \omega^1q^0&:& y_{10}''+\frac{i}{1-u}y_{00}'+\left[\frac{(uf)'}{uf}-c\right]y_{10}' +\left[\frac{i}{2(1-u)^2}+\frac{i((uf)'-cuf)}{2uf(1-u)}\right]y_{00} \non
&&\hspace{4in} -c\frac{(uf)'}{uf}y_{10}=0, \nonumber \\
 \omega^0q^2&:& y_{02}''+\left[\frac{(uf)'}{uf}-c\right]y_{02}'-c\frac{(uf)'}{uf}y_{02}-\frac{f}{uf^2}y_{00}=0.
\end{eqnarray}
Using the same analysis for the low frequency of spectral function as
described in the previous Appendix, the solutions regular at $u=1$ 
are found to be
\begin{eqnarray}
 y_{00}&=&A e^{cu} \nonumber \\
 y_{10}&=&\frac{iA}{2}e^{cu+c}\left[C_{10}+2\mbox{Ei}(-cu)-e^c\mbox{Ei}(-c(1+u))-e^{-c}\left(\mbox{Ei}(c(1-u))-\ln(u-1)\right)\right]\nonumber\\
y_{02}&=&\frac{A}{2c}e^{cu+c}\left[C_{02}-2\mbox{Ei}(-cu)+e^c\mbox{Ei}(-c(1+u))
\right. \nonumber\\ &&\left.+
e^{-c}\left(\mbox{Ei}(c(1-u))+2\ln u-\ln(u^2-1)\right)
\right],
\end{eqnarray}
where $A$ and $C_{10},C_{02}$ are constants independent of $u$. We can determine $A$ in terms of the boundary values of $V_t$ and $V_{\parallel}$ at $u\to 0$ defined as
\begin{eqnarray}
 \lim_{u\to 0} V_t(u)&=&V_t^0,\nonumber\\ 
 \lim_{u\to 0} V_{\parallel}(u)&=&V_{\parallel}^0.
\end{eqnarray}
Substituting the solution for $E=V_t'$ into equation (\ref{Vz in terms
  of V_t}) and taking limit $u\to 0$, the integration constants $C_{10}, C_{02}$ drop out and we can determine $A$ to be
\begin{equation}
 A=\frac{q^2V_t^0+\omega q V_{\parallel}^0}{i\omega e^c-\frac{e^c}{c}\left(1-e^{-c}\right)q^2+O(\omega^2,\omega q^2,q^4)}.
\end{equation}
We recognize the hydrodynamic pole and as explained in
\cite{policastro02} we can now compute the time-time component of the retarded thermal Green's
function of two currents
\begin{equation}
 G_{tt}=\frac{N_c^2 T^2 q^2 e^{-c}}{8(i\omega-\frac{\left(1-e^{-c}\right)}{c}q^2)}+\cdots,
\end{equation}
Thus the time-time component
of the spectral density function at low frequency and momentum equals
\begin{equation}
 \chi_{tt}(k^0,\vec{k})=-2~\Im[G_{tt}]=\frac{N_c^2 T k^0 |\vec{k}|^2 e^{-c}}{8\pi((k^0)^2+D |\vec{k}|^2)}+\ldots,
\end{equation}
with $D=\frac{\left(1-e^{-c}\right)}{2\pi T c}$ the diffusion
constant. Comparing the result with the universal hydrodynamic
behaviour
\begin{equation}
 \chi_{tt}(k^0,\vec{k}) =\frac{2\omega D |\vec{k}|^2}{(k^0)^2+(D|\vec{k}|^2)^2}\Xi +\dots,
\end{equation}
the charge susceptibility $\Xi$ is seen to equal $\Xi=\frac{N_c^2 T^2 c}{8(e^c-1)}$ and naturally satisfies the Einstein relation $\Xi=\sig/e^2D$.

\end{document}